# Transport and Photo-Conduction in Carbon Nanotube Fibers

**O. S. Dewey, R. J. Headrick, L. W. Taylor, and M. Pasquali**

*Department of Chemical and Biomolecular Engineering*

*Rice University, Houston, Texas 77005, USA*

**G. Prestopino and G. Verona Rinati**

*Dipartimento di Ingegneria Industriale, Università di Roma "Tor Vergata", 00133 Roma, Italy*

**M. Lucci and M. Cirillo***

*Dipartimento di Fisica and Minas Lab, Università di Roma "Tor Vergata", 00133 Roma, Italy*

## Abstract

We have characterized the conductivity of carbon nanotubes (CNT) fibers enriched in semiconducting species as a function of temperature and pulsed laser irradiation of *266 nm* wavelength. While at high temperatures the response approaches an Arrhenius law behavior, from room temperature down to *4.2 K* the response can be framed, quantitatively, within the predictions of the fluctuation induced tunneling which occurs between the inner fibrils (bundles) of the samples and/or the elementary CNTs constituting the fibers. Laser irradiation induces an enhancement of the conductivity, and analysis of the resulting data confirms the (exponential) dependence of the potential barrier upon temperature as expected from the fluctuation induced tunneling model. A thermal map of the experimental configuration consisting of laser-irradiated fibers is also obtained via COMSOL simulations in order to rule out bare heating phenomena as the background of our experiments.

(*) Author to whom correspondence should be addressed : cirillo@roma2.infn.it

Carbon nanotube (CNT) fibers represent one of the most promising and reliable carbon allotrope materials synthesized to date [1,2,3], and we address the reader to ref. 4 where a review covering both fundamental aspects and applications can be found. The properties of the fibers significantly enhance the variety of possible applications of CNTs which include devices operating in high frequency regions of the electromagnetic (EM) spectrum [5,6]. Very recently, CNT fibers engineered in a form of diodes, have shown the possibility of being used as textile materials sensitive to elementary particle and/or EM radiation [7]. The versatility of the fibers to be doped [8] or coated [9], opens up a variety of possibilities for tuning their functional behavior generating what are considered as metamaterials and metadevices.

We herein concentrate our attention on characterizing the electrical and photoconductive response of CNT fibers that were engineered to have a significant fraction of semiconducting nanotubes as elementary constituents. In the high temperature limit, we estimate the gap value from an Arrhenius-like dependence [10-12] but the fluctuation induced tunneling (FIT) [13-16] of charge carriers provides a reasonable account of the data over the investigated temperature range with and without laser light irradiation.

CNT fibers were prepared using a solution processing method described previously [1]. CNTs enriched in *(6,5)* chirality (CoMoCAT, SG65i, batchPXD1-13) were added to chlorosulfonic acid and mixed on a stir plate overnight. This solution was loaded into a stainless steel syringe and extruded through a spinneret into a coagulation bath. The fibers were collected onto a rotating drum, washed in room temperature water, and dried in an oven at *115 °C* for *12* hours. The dried fibers were then washed in water at *60 °C* for 3 hours and allowed to dry at ambient conditions.

In Fig. 1a,b,c,d we show scanning electron spectroscopy (SEM) images (FEI Helios NanoLab 660) of a fabricated fiber with increasing resolution with (a) the lowest magnification and (d) the highest; the inner structure of the fibers consist of fibrils (bundles) with diameters in the range of *5-25 nm*. These bundles are evident in (c) and (d). We can see that contact among the bundles may take place either between the lateral surfaces of aligned bundles or by the lateral branches. Resonant Raman spectroscopy was obtained on the CNT fibers with a Renishaw inVia confocal microscope at *633 nm* and *532 nm* excitation wavelengths. In Fig. 2a we show typical spectra. The G/D band ratio for each spectrum is measured to be ~*4*. The CNT diameters are estimated from the radial breathing modes (RBM) using the relation $\omega_{RBM} = \frac{218.3}{d_t} + 15.9$ where $\omega_{RBM}$ is the wavenumber of the RBM in $cm^{-1}$ and $d_t$ is the diameter of the CNT in *nm*. This relation has been shown to be accurate for small diameter CNTs [3]. For each excitation wavelength, we use the Kataura plot to highlight metallic and semiconducting regions in orange and purple, respectively.

To obtain a more comprehensive estimate of the diameter distribution absorbance spectroscopy was obtained with a Shimadzu UV-vis spectrometer and the results are shown in Fig. 2b. The sample was dispersed in 1 wt/v% sodium deoxycholate surfactant solution (DOC) by 2 hours of tip sonication at ~*0.1 mg/mL*. The sample was then centrifuged at *13,000* rpm for 1 hour and diluted with DOC prior to taking the absorbance spectra. The strongest peak in the spectra occurs at about *980 nm* and is resultant from the $S_{11}$ electronic transition of the *(6,5)* CNT indicating that the sample is enriched in *(6,5)* chirality [17], as is common for SG65i CoMoCAT CNTs. The *(6,5)* chirality CNT has a diameter near *0.76 nm,* but its radial breathing mode in the Raman spectra is very weak compared to the other CNTs that are present in the sample despite the fact that the sample is enriched in *(6,5)* chirality. This is due to a lack of resonance with the excitation wavelengths and an inherently weak RBM phonon mode for the *(6,5)* chirality [18].

Fig. 3 displays the schematic of our experimental setup. The fibers were secured by silver paste onto gold electrodes that were previously deposited on an oxidized silicon wafer. All the resistance measurements were performed according to a standard four-point probe configuration. These measurements, down to *4K,* were performed keeping the samples in a liquid helium bath, whereas the experiments under laser irradiation were performed in a Gifford-McMahon cryocooler with an optical quartz window through which the laser light could couple to the sample, see Fig. 3.

A Q-switched Nd YAG laser with a fourth harmonic generator was used, providing *2 ns* pulses with a repetition frequency of *10 kHz* and a *266 nm* wavelength. The cold finger of the cryocooler could reach temperatures down to *40 K* and a heater in contact with the cold finger enabled temperatures up to *400 K*. The results presented here are very representative of the investigated samples. It is worth noting that we also performed the same type of experiments using a He-Ne *632.8 nm* wavelength laser, and the results we observed in this case were qualitatively similar to those obtained with the *266 nm* wavelength laser. A report on the results obtained with lasers of different wavelengths will be dealt with in a future publication.

In Fig. 4a we show a typical temperature dependence of the resistance of one fiber measured in the range (*4K-40K*), while the inset shows the dependence in the whole investigated temperature range (*4 K-300 K*). The curves in the plot were obtained for different bias currents ranging from *100 nA to 500 μA*. In Fig. 4b we show an Arrhenius-like dependence (resistance vs *1/T*) for a bias current of *100 nA*. This dependence is shown in detail in the inset of Fig. 4b. Deriving energy gap data from the high temperature resistance Arrhenius dependencies is a custom experimental protocol [10-12]. From the plot we extract a slope of (*73± 2) K*. Assuming that this number equals the modulus $E_g/2k_B$ and given the value of the Boltzmann constant *($k_B$=8.616 x 10$^{-5}$ eV/K)*, we calculate an energy gap of $E_g$ = *12.6 meV*. All of the other data shown in Fig. 4a, obtained for different currents, exhibited the same behavior at high temperatures and had nearly the same energy gap value.

We employ now the functional dependencies provided by the FIT model [13-16] to follow the *R(T)* dependence over the whole temperature range investigated. The physical background of the model is that conduction is generated through tunneling across barriers between CNT bundles or CNTs, and the temperature dependence is due to variations of the tunneling energy across the barriers. According to the model, the dependence of the resistance on the temperature is $R/R_0 = e^{\frac{T_1}{(T+T_0)}}$ where $R_0$ is the room temperature resistance of the sample while $T_0 = \frac{4\hbar S V_0^{3/2}}{\pi^2 w^2 k_B e^2 \sqrt{2m}}$ and $T_1 = \frac{2 S V_0^2}{\pi w k_B e^2}$.

In these two expressions $V_0$ is the depth of the potential well (in other terms an energy gap) which develops across barriers physically defined by a surface $S$ and a width $w$; the constants $e$ and $m$ are electron charge and mass respectively, and $h = 4.136\ 10^{-15}\ eV\cdot s$ is the Planck constant where $\hbar=h/2\pi$. The results of the fitting procedures are shown by the curves through the experimental data in Fig. 5a. The values of $T_0$ and $T_1$ for the three best-fit curves are $T_0= 10.41\ K$ and $T_1= 33.02\ K$ for the *1 μA* bias, $T_0=12.1K$ and $T_1=33.38K$ for *100 μA* bias, and $T_0=20.0K$ and $T_1=39.5K$ for the *500 μA* bias with standard fit errors of a few percent over these values.

In Fig. 5b the plot used to determine the height of the potential $V_0$ in the limit of low current and low temperature is reported. The evaluation goes through the ratio $\frac{T_1}{T_0} = \frac{\pi w \sqrt{2mV_0}}{2\hbar}$, calculated for all the bias currents of Fig. 2. From this plot and from an estimate of the value of $w$ (of the order of tenths of nanometers [15,19]), we conclude that $V_0 \cong 1.1\ eV$. Thus, the depth of the well in the low current and low temperature region, according to the FIT model, is roughly two orders of magnitude larger than the one obtained from the Arrhenius plot approximation at high temperatures.

In general, the potential of the barrier $V_0$ in the FIT model depends on *I* and *T*. Assuming that the two dependencies can be separated to give $V_0 = V_1(T)\ ln^2 I$, the dependence of $V_1$ upon temperature is expected to be exponential [15]. We obtain evidence of this dependence from the characterization of the resistance of the fibers when irradiated with laser light of *266 nm* wavelength. The results are shown in Fig. 6a: the two datasets displayed in the curves demonstrate the temperature dependence of the resistivity with and without laser irradiation. The respective arrows indicate how temperature was varied during each measurement. The effect we measure is reminiscent of that observed for CNTs deposited over multifingers and conditioned by RF fields [20].

The effect of laser irradiation is to supply an energy $h\nu$ (on the order of *4.66 eV* in the present case, considering the *266 nm* wavelength) which in turn flips charge carriers in the conduction band with a consequent increase of the conductivity (or decrease of the resistivity). At higher temperatures the energy input to the charge carriers by thermal excitations is already high enough to provide a flipping over the gap (as the Arrhenius fit shows). At lower temperatures the thermal excitations have a lower energy, so the laser induces greater tunneling as compared to the high temperature case. In Fig. 6b we show a semi-log plot of the resistance variation upon laser irradiation as a function of temperature extracted from the plot of Fig. 6a, i. e. the difference between the measured resistance values between the laser-on and laser-off cases . The slope obtained was the same for other investigated samples. The solid line we see through the data in Fig. 6b is an exponential fit of the data with a function *y= A exp(B/T)* where *A=(503±3) Ω* and *B=-0.0126 K$^{-1}$* . The excellent fit of the data obtained with an exponential law confirms the behavior predicted by the FIT model [15]. Extrapolating the data to low temperature (as mentioned previously, our cryocooler equipped with the optical window could only reach temperatures of about *40 K*), we can see that the variation roughly covers the two orders of magnitude difference that we have obtained for the energy gap from the Arrhenius approximation at high temperature (*12.5 meV*) and from the FIT model at low temperatures (about *1eV*).

The measured low temperature limit of the energy gap of our samples is also consistent with the predictions from a tight binding approximation for CNT systems considering the diameter of the elementary CNTs [21-23]. In this physical limit case, the value of the energy gap is given by $E_G=\beta/D$, where *β≈0.7 eV·nm*. Considering that our CNTs diameter of *0.76 nm* an energy gap of the order of *1eV* can be expected. We speculate that the highest values of the gap are set by the ultimate the aligned semiconducting CNTs densely packed inside the bundles [1].

In order to check possible thermal effects on our laser irradiation experiments, we have set up a full COMSOL [24] simulation taking into account all the physical details of our experimental configuration. This has allowed us to obtain a thermal map of the region of interest containing the CNT fiber sample upon irradiation. In Fig. 7 we can see the schematic setup showing the fiber secured by silver paste (represented by the parallelepiped over the gold contacts) to the gold electrodes. The gold electrodes lay on *1μm* layer of *SiO₂* , have a width of *1 mm* and their centers are *2.54 mm* apart. When the *10mW* laser radiation laser reaches the sample, the beam is roughly distributed over an area of (*1mm x 2mm)* and a power of about *250μW* reaches the fiber (considering an average fiber diameter of *25μm* diameter). The thermal conductivities of the different materials are set by the software. For the thermal conductivity of the fibers we have imposed a value of *500 W/mK* as reported previously [1]. The temperature scale on the right shows the increase in temperature of the fiber caused by the laser radiation. The order of magnitude of temperature increase in the region where the laser is directed is a few tenths of a Kelvin while we estimate that the temperature input necessary to generate a change in resistance like that shown in Fig. 6a would be of the order of several tens of Kelvin. Thus, heating generated by the laser light on the fibers has a negligible effect on our experiments and the only effect of the light is to influence the charge activation process. The thermal contact of the fiber with the gold pads, secured by the silver paste, contributes substantially to limit the raise in temperature.

We attempted to compare our data with the variable hopping range (VHR) model as used in ref. 25, but we could not get a fit of our data. We suppose that the characteristics of our highly aligned CNT fibers are rather different [26,27] than the CNT buckypaper structure investigated in that specific paper. The FIT also provides motivation for the work presented in ref. 28 treating CNT networks as disordered percolating systems. However, in ref. 28 the analysis is supposed to be valid at temperatures above *90 K*. As far as the alignment of the tubes is concerned the CNT samples of ref. 29 would be somewhat closer to our experimental configuration, but in this reference the specific dependence of the transport properties upon temperature is not investigated, due to the focus on transistor applications. Framing the data with the FIT model is also an indirect evidence of the concerns of ref. 30 on polymers generating potential barriers between CNTs.

In conclusion, we have analyzed the temperature dependence of the transport properties of CNT fibers fabricated from CNTs of semiconducting nature. Between the two limits of high temperature (Arrhenius-like) and low temperature (consistent with a tight binding expectation) the data follow the predictions of the FIT model. Considered the variations of voltage generated by the laser irradiation on the fibers, we speculate that either the bare fibers or their adequate functionalized products could lead to practical detectors of EM radiation or elementary particles.

**FIGURE CAPTIONS**

**Figure 1**. SEM photos of a typical CNT fiber that we have investigated at increasing magnification: (a), (b), (c), and (d). The black boxes indicate the portion to be magnified. Images (c) and (d) clearly depict the tightly-packed fibril (bundles) structure and the close contacts between these. Each bundle, in turn, was originated by aligned nanotubes having an average diameter of *0.76 nm*.

**Figure 2**. (a) Resonant Raman spectra was obtained on the CNT fibers with a Renishaw inVia confocal microscope at *633 nm* and *532 nm* excitation wavelengths; (b) absorbance spectroscopy obtained with a Shimadzu UV-vis spectrometer. The spectrum indicates that the sample is enriched in *(6,5)* chirality.

**Figure 3**. Sketch (not to scale) of our experimental setup. The light from a laser source is coupled to a CNT fiber mounted on the cold finger of a cryocooler. Gold contacts evaporated on an oxidized silicon chip provide access to DC four probe measurements.

**Figure 4**. (a) The temperature dependence of the resistance of one fiber obtained for different values of the bias current (inset shows the behavior up to *300K*). (b) One of the curves shown in (a) (bias current *I=100nA*) plotted in an Arrhenius plot; the inset is an enlargement of the area in the rectangle demonstrating the straight line behavior at high temperatures.

**Figure 5**. (a) Data from Fig. 4a displayed in a semi-log plot as a function of *1/T* and fitted in terms of the fluctuation induced tunneling (FIT) model (solid lines); (b) the plot used to determine the potential depth of the FIT model at low temperatures.

**Figure 6**. (a) The variation in resistance of the fibers generated by laser irradiation. The lower resistance curve upon laser irradiation is indicated in the figure. The two slightly different curves that we see for the laser off case were obtained by cooling or heating the sample (as indicated by the respective arrows). (b) The difference in resistance between the two curves

in (a), namely the "laser off" ($R_0$) and "laser on" ($R_L$) plotted against temperature in a semi-log plot demostrates a straight exponential behavior of the resistance decrease.

**Figure 7**. Sketch of the basis for our COMSOL simulation for the laser-irradiated fibers. The scale on the right-hand side of the picture shows the temperature scales in terms of color. We can see that the temperature of the fiber affected by the laser radiation rises only of a few tenths of a degree Kelvin. The gold contact pads *P1, P2, P3, P4* are *1mm* wide and the spacing between their centers is *2.53 mm*.

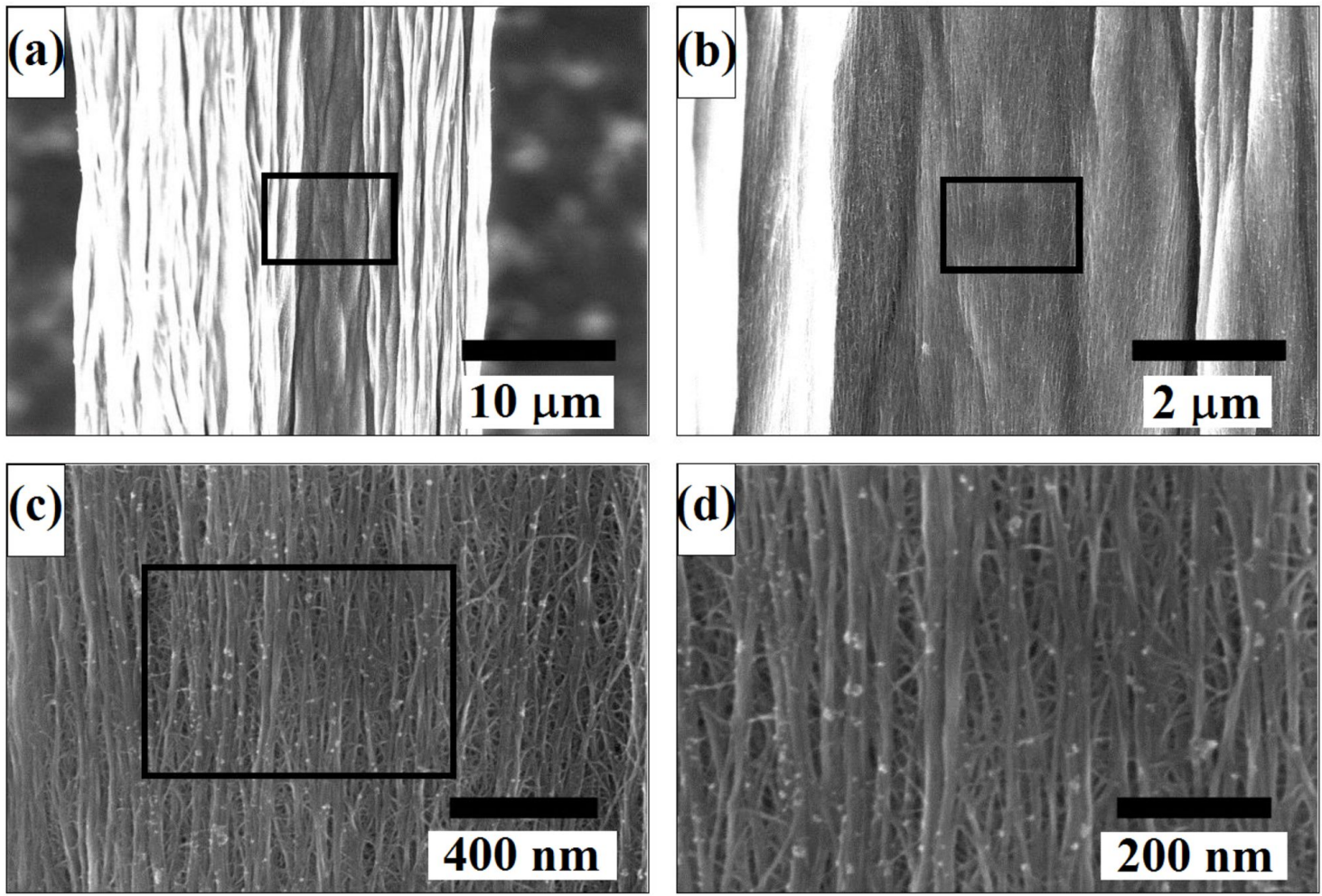

O. S. Dewey et al., Fig. 1

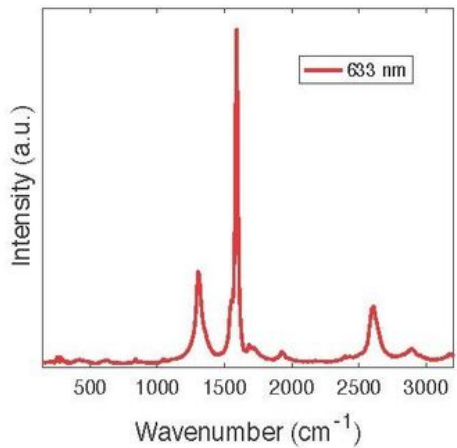 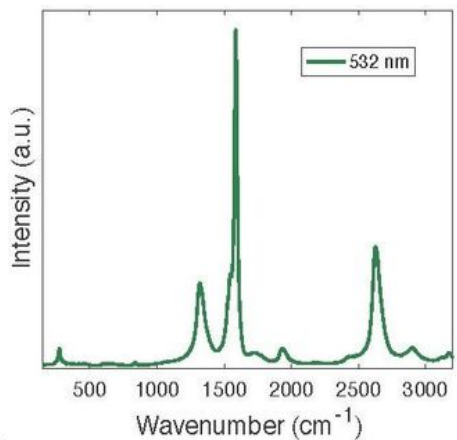

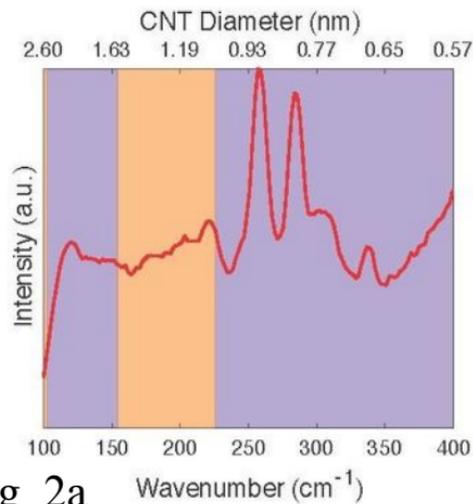 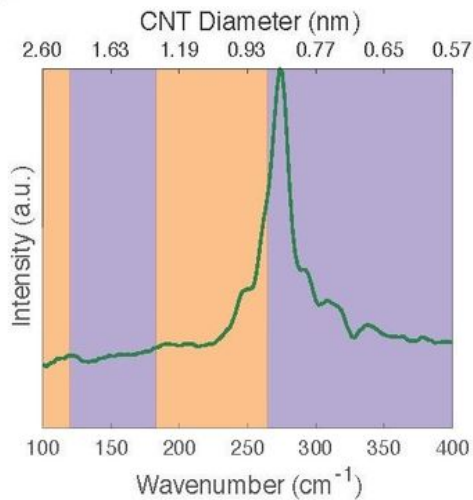

Fig. 2a

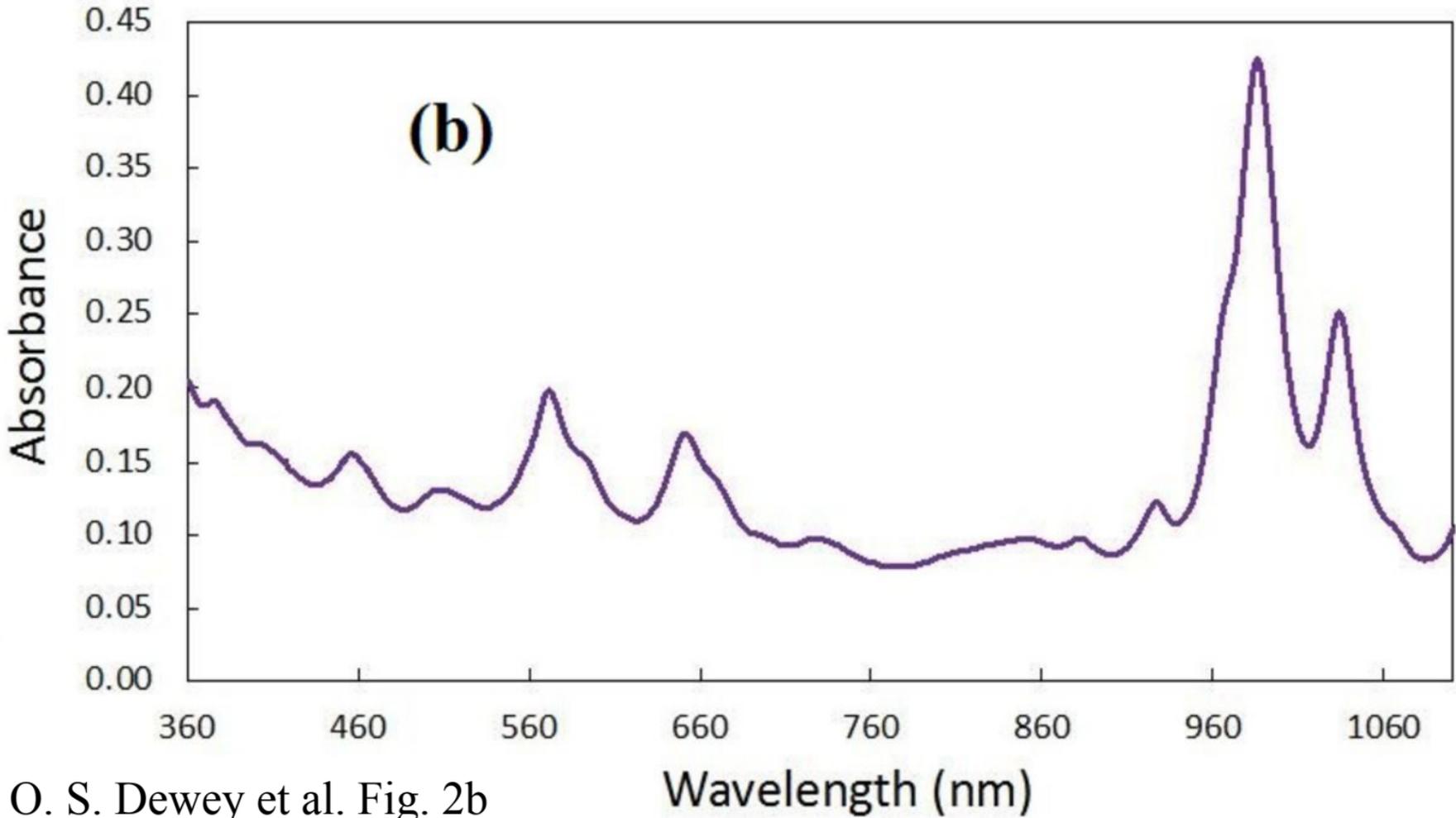

O. S. Dewey et al. Fig. 2b

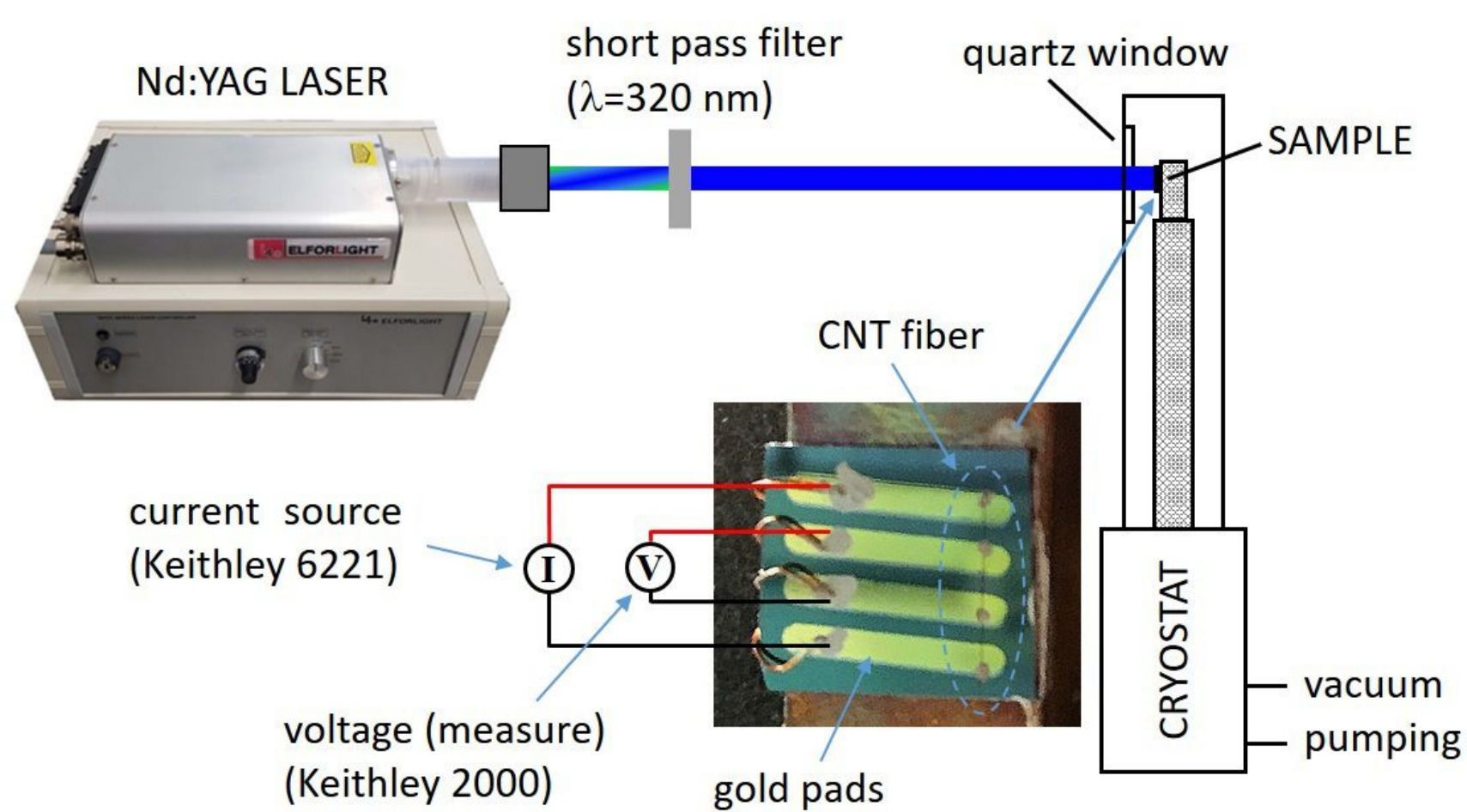

O. S. Dewey et al. Fig. 3

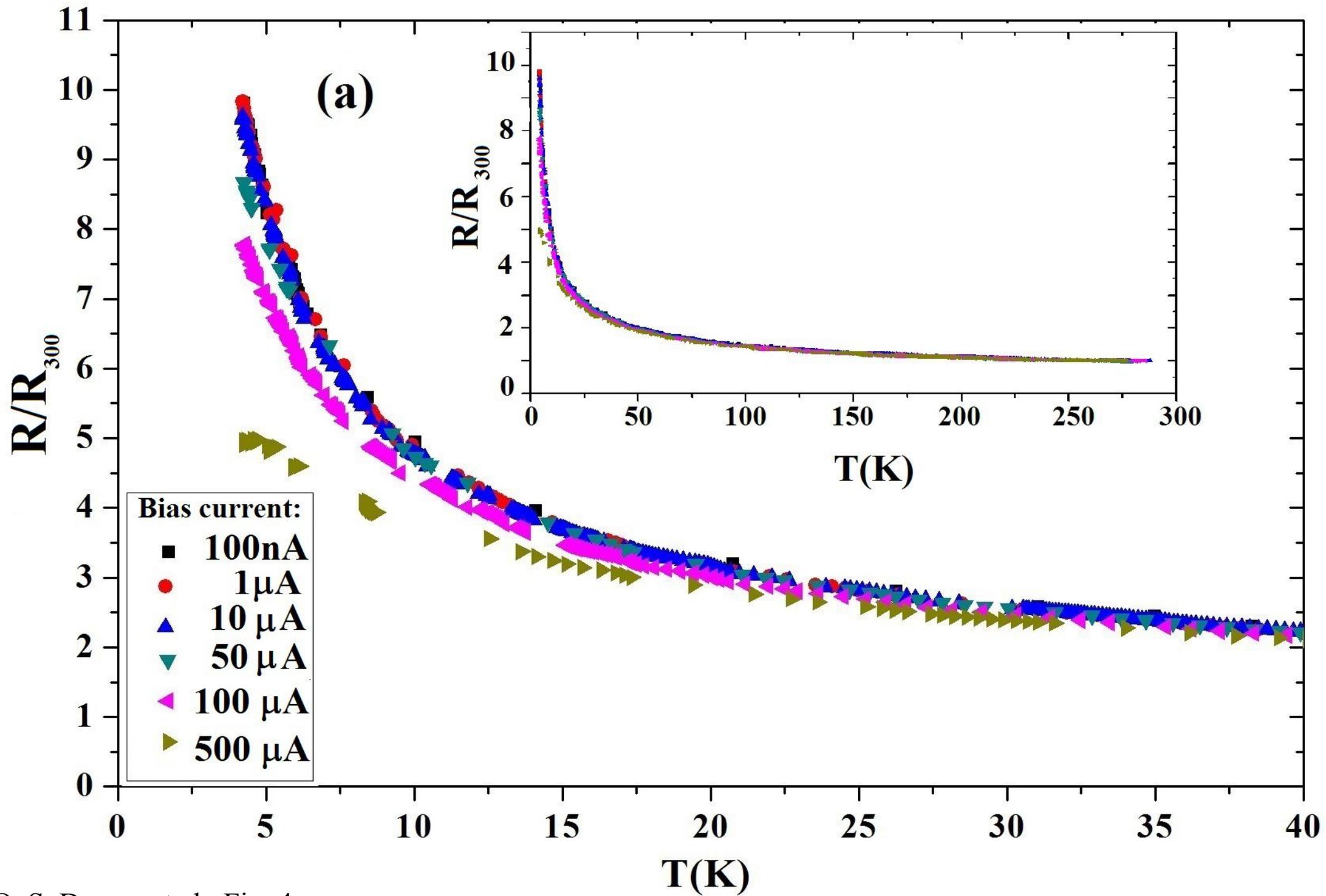

O. S. Dewey et al., Fig. 4a

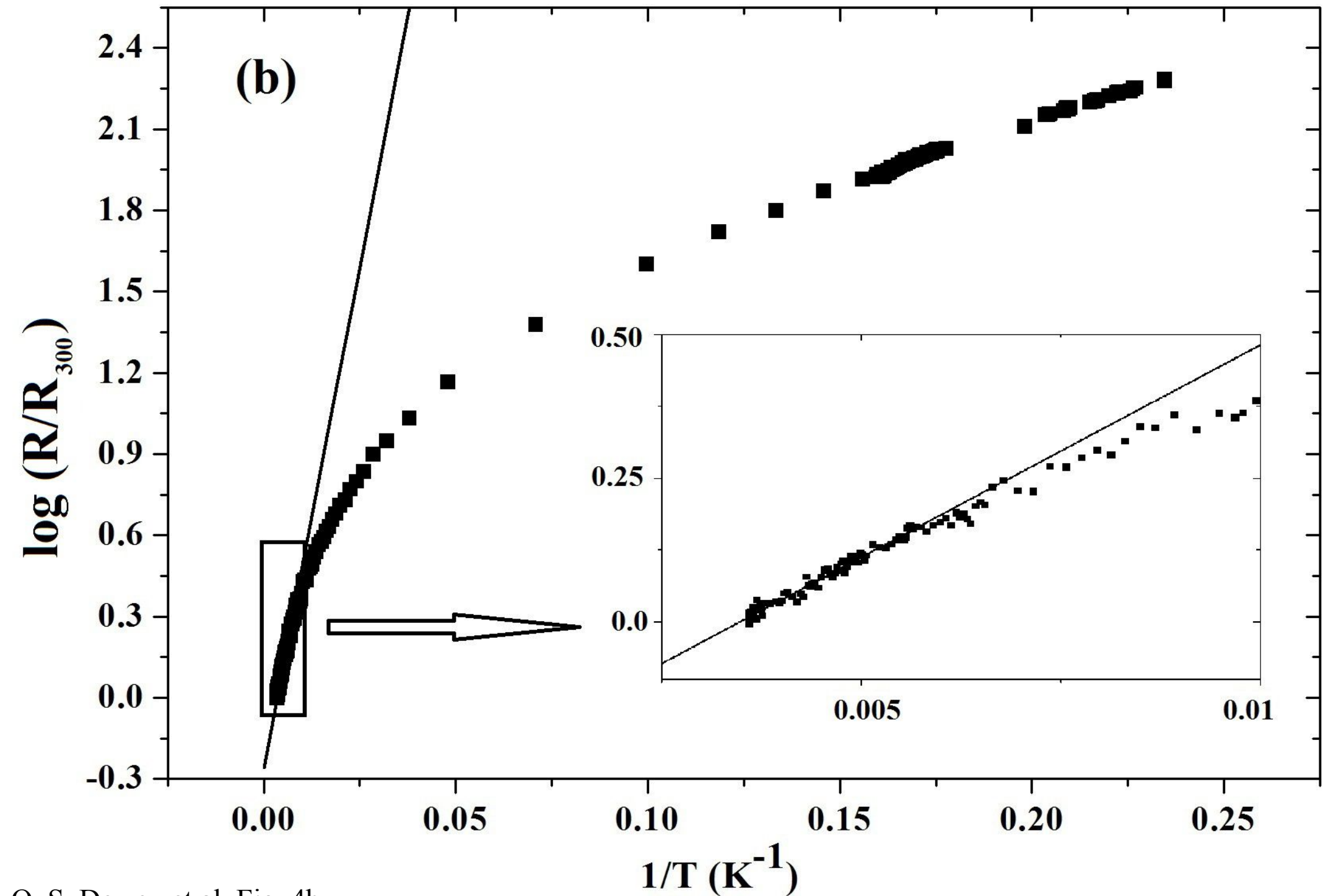

O. S. Dewey et al. Fig. 4b

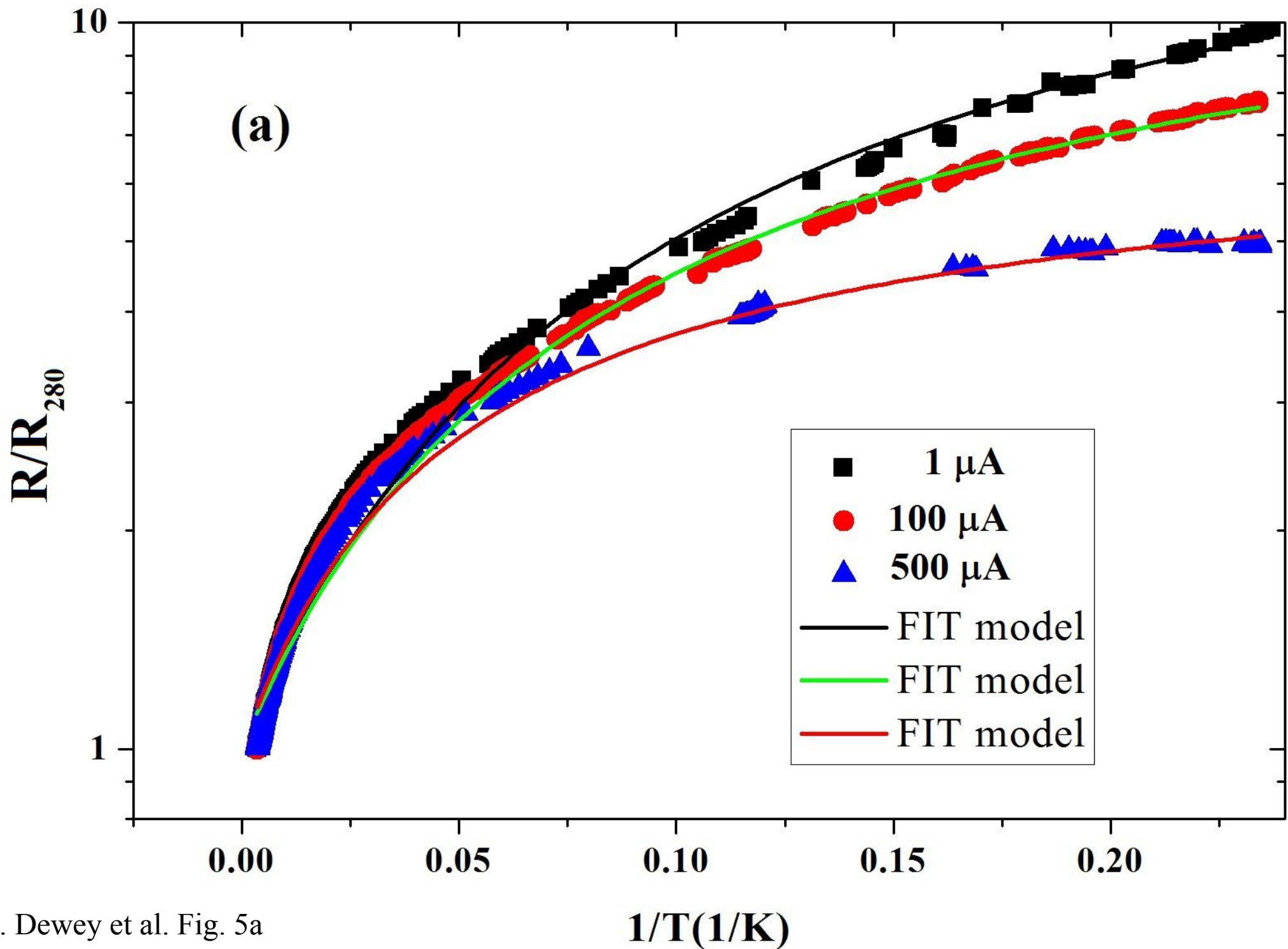

O. S. Dewey et al. Fig. 5a

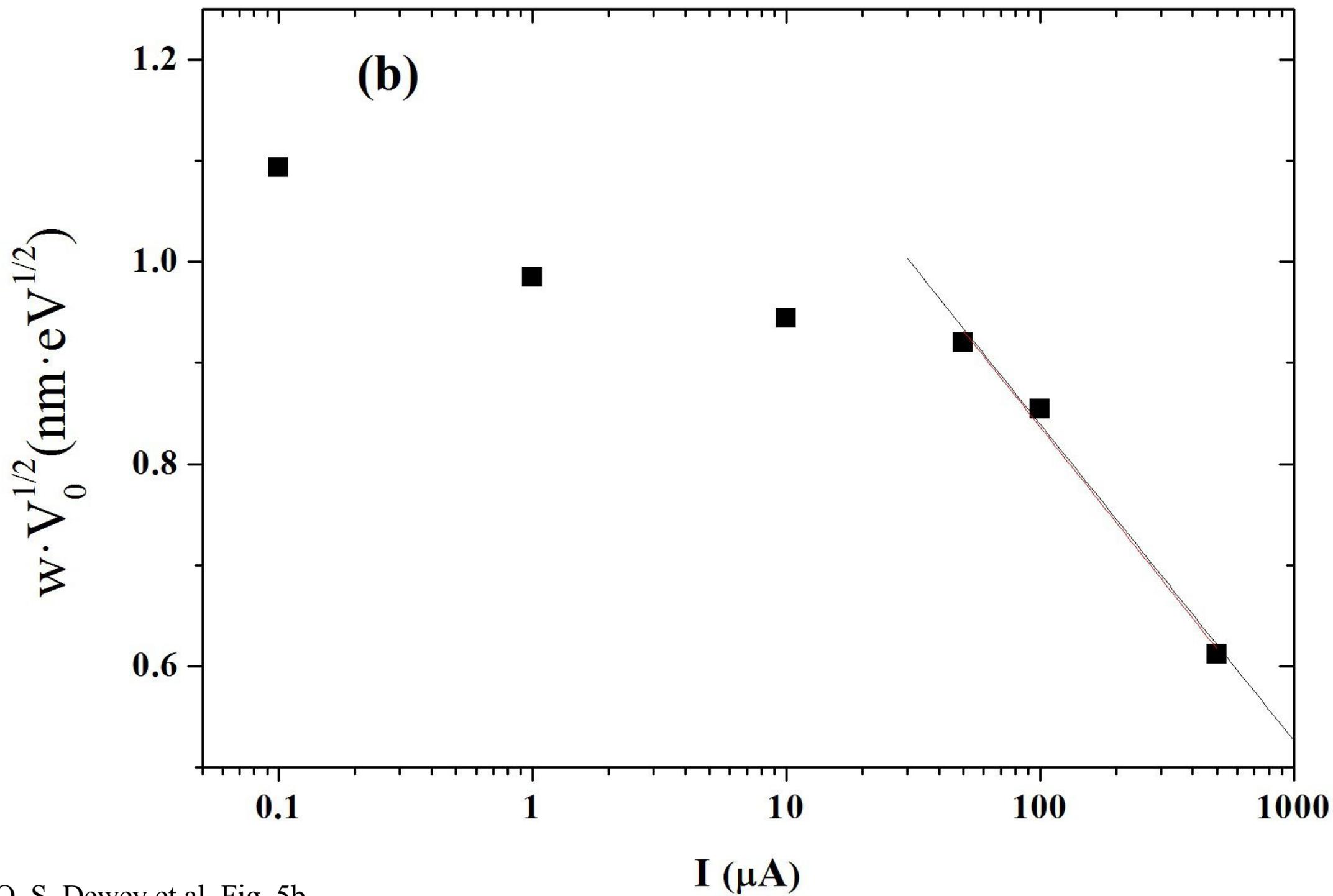

O. S. Dewey et al. Fig. 5b

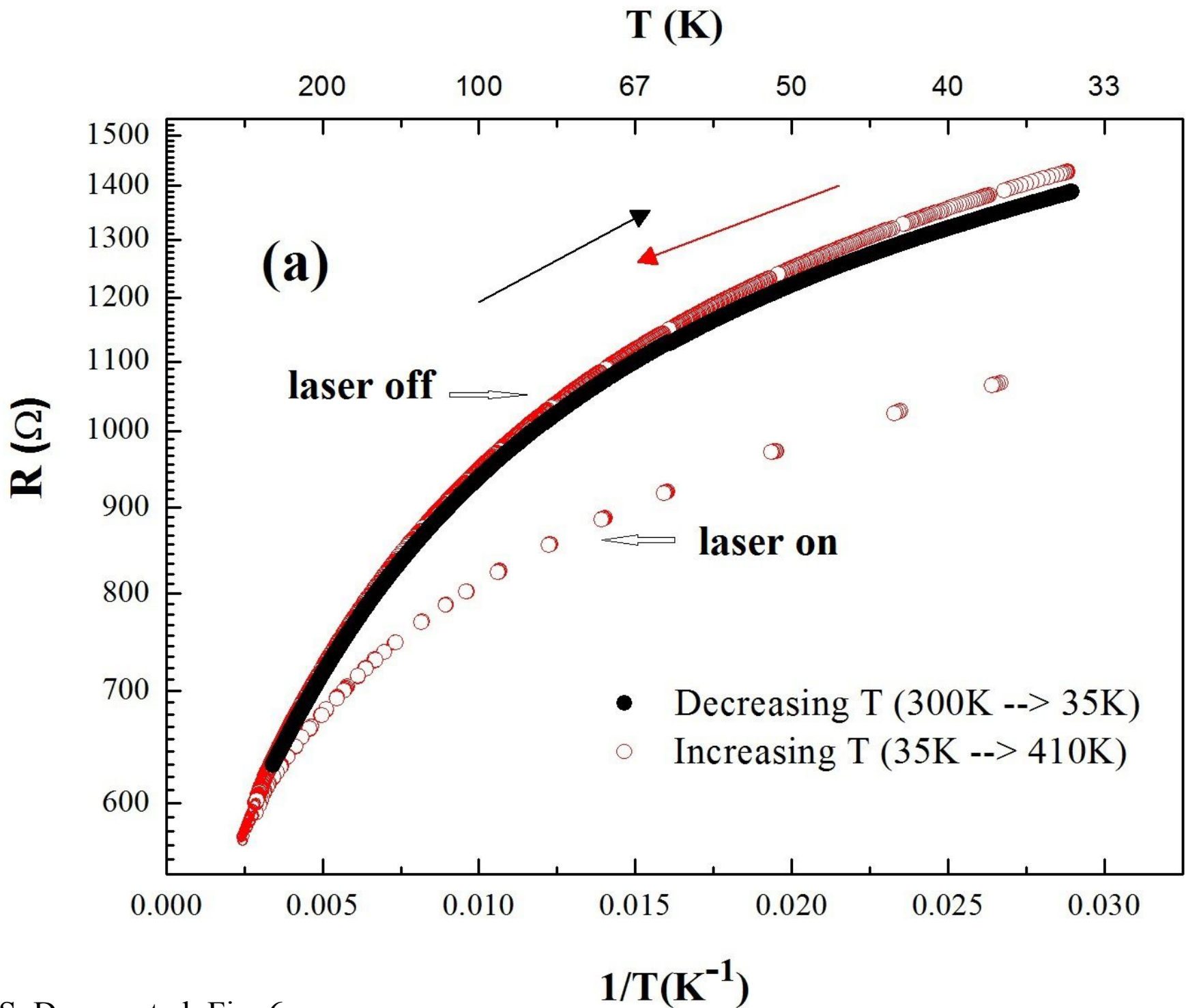

O. S. Dewey et al. Fig. 6a

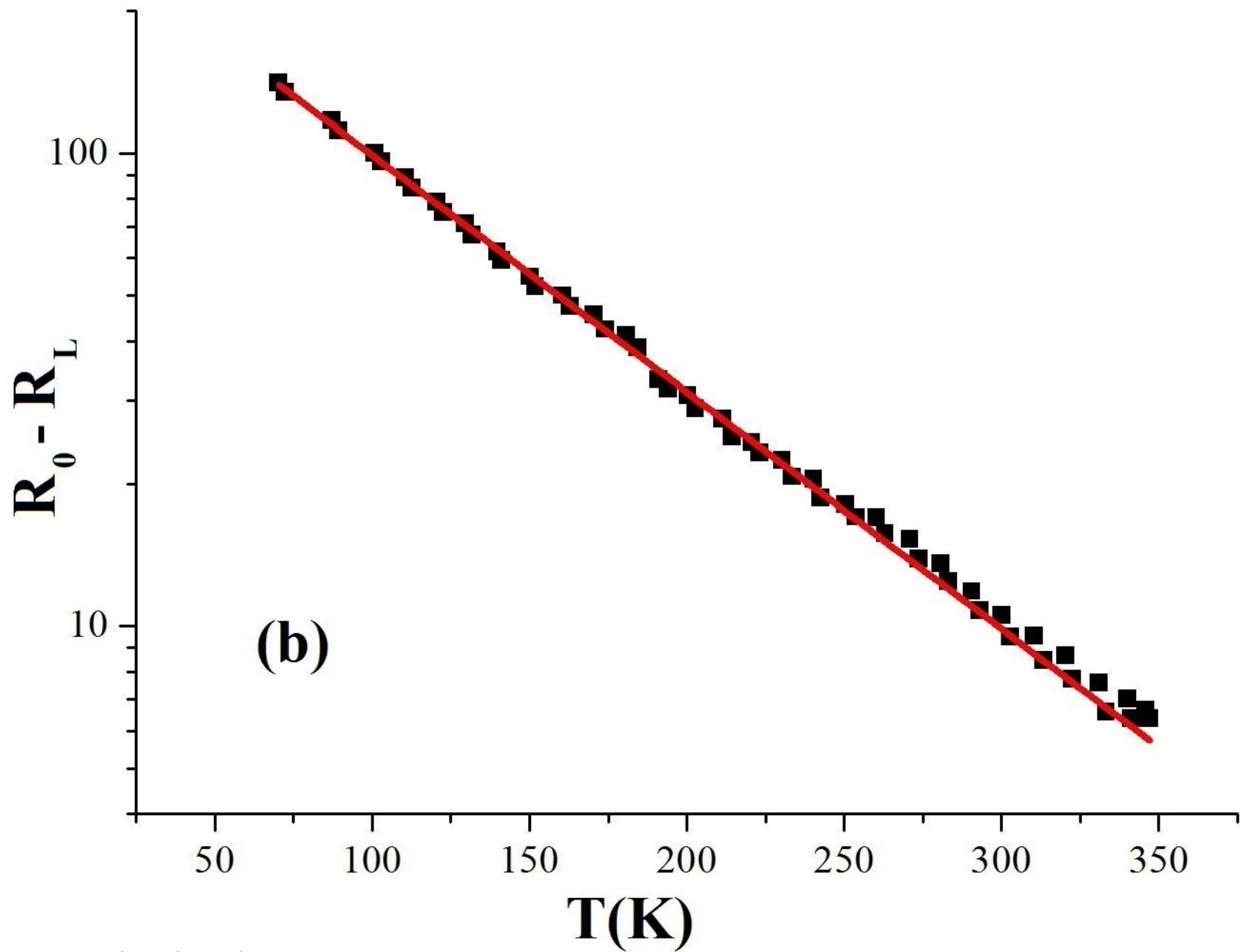

O. S. Dewey et al., Fig. 6b

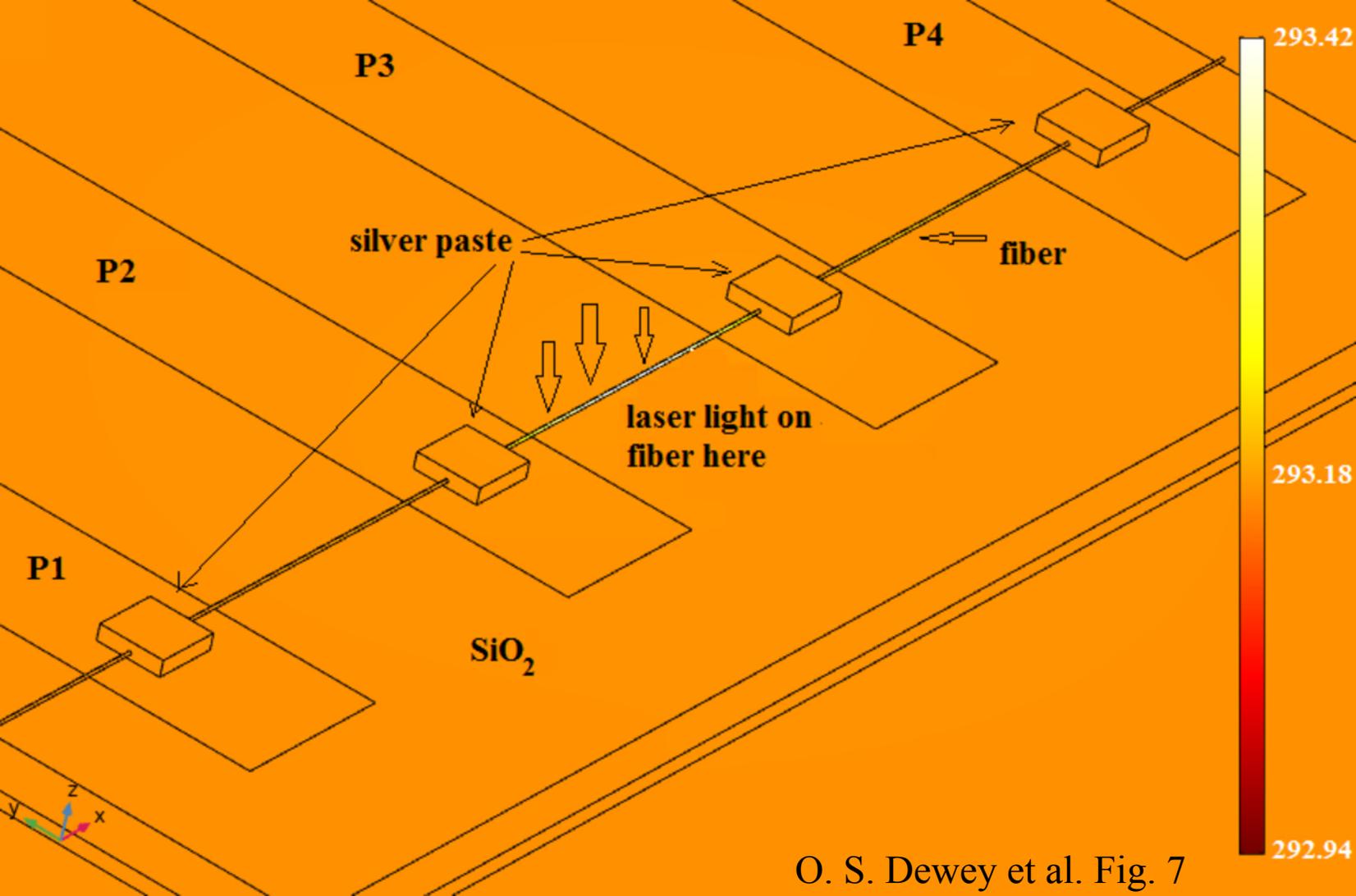

O. S. Dewey et al. Fig. 7